\documentclass[aas2pp4,epsfig]{aastex}
\newcommand\cyg{Cygnus~X-3}
\newcommand\solar{$M_\odot$}

\shorttitle{VLBA Images of Cygnus X-3}
\shortauthors{Mioduszewski et~al.}
\slugcomment{accepted for publication in ApJ}

\begin{document}
\renewcommand\topfraction{.9}
\renewcommand\dbltopfraction{.9}

\title{A One-sided, Highly Relativistic Jet from Cygnus X-3}
\author{Amy J. Mioduszewski\altaffilmark{1,2}}
\affil{RCfTA, School of Physics, University of Sydney, NSW 2006, Australia}
\email{amiodusz@physics.usyd.edu.au}
\author{Michael P. Rupen, Robert M. Hjellming\altaffilmark{3}}
\affil{NRAO, P.O. Box 0, Socorro, NM 87801, USA}
\email{mrupen@nrao.edu}
\author{Guy G. Pooley}
\affil{Mullard Radio Astronomy Observatory, Cavendish Laboratory,
Madingley Road, Cambridge CB3 OHE, England}
\email{ggp1@cam.ac.uk}
\author{Elizabeth B. Waltman}
\affil{Remote Sensing Division, Naval Research Laboratory, Washington
DC 20375, USA}
\email{ewaltman@rsd.nrl.navy.mil}
\altaffiltext{1}{JIVE, Postbus 2, 7990 AA, Dwingeloo, the Netherlands}
\altaffiltext{2}{present address: NRAO, P.O. Box 0, Socorro, NM 87801, USA}
\altaffiltext{3}{deceased, July 29, 2000}

\begin{abstract}
Very Long Baseline Array images of the X-ray binary, \cyg\, were
obtained 2, 4 and 7 days after the peak of a 10~Jy flare on 4
February 1997.  The first two images show a {\it curved one-sided} jet,
the third a scatter-broadened disc, presumably at the position of the core.
The jet curvature changes from the first to the second epoch, which strongly
suggests a precessing jet.  The ratio of the flux density in the approaching
to that in the (undetected) receding jet is $\gtrsim330$; if this asymmetry
is due to Doppler boosting, the implied jet speed is $\gtrsim 0.81c$. 
Precessing jet model fits, together with the assumptions that the jet is
intrinsically symmetric and was ejected during or after the major flare,
yield the following constraints:
  the jet inclination to the line of sight must be $\lesssim14^\circ$;
  the cone opening angle must be $\lesssim12^\circ$; 
  and the precession period must be $\gtrsim60\rm\,days$.
\end{abstract}


\section{Introduction}

\cyg\ is one of the few X-ray binaries that is consistently strong in both
radio and X-ray emission.  In fact, in both quiescent and flare states, \cyg\ 
is the most luminous X-ray binary at radio wavelengths \citep{Walt95}.
Its companion is thought to be a Wolf-Rayet star \citep{vKetal92},
and a 4.8~h period has been seen at X-ray (e.g., \citet{MS79})
and infrared \citep{MCW86} energies.
This period is generally interpreted as the orbital period of the binary
system.  \citet{SGS96} used the velocity shifts of lines
in the infrared to estimate, assuming the orbital system is seen edge-on, that 
the compact object is a black hole with mass~$\gtrsim 7$~\solar.

  Giant radio outbursts have been known in \cyg\ since 1972 \citep{Getal72}. 
Large radio flares are always preceded by a quenching of the radio
emission \citep{Walt94}.  The hard X-ray is usually anti-correlated
with the radio except during quench-flare periods, although the quench period
in hard X-ray is generally longer than in the radio \citep{Metal99}.
Similarly, the soft thermal X-ray and radio emission is usually 
correlated except during quench-flare periods \citep{Wat94}.
In other words, before a large radio/hard X-ray flare
the soft X-ray is more luminous than usual, and during a flare the soft X-ray
intensity briefly drops.  Observations
with the
VLA \citep{Getal83}, MERLIN \citep{SSJH86}, and VLBI
\citep{Setal95, MRG88, Setal99}, 
during or shortly after
large radio flares, suggest ejection of radio-emitting plasma in the
north-south direction.   \citet{SSJH86}, \citet{Setal95},
and \citet{MRG88}, assuming a two-sided jet, estimated an
expansion rate of {\rm $\sim$ 5~$\rm mas/day$.
This corresponds to an apparent transverse speed of
$0.3c$ for a distance of 10~kpc, which we adopt for the remainder of the paper
(based on \ion{H}{1} absorption;
see Dickey 1983).  However note a recent \ion{H}{1} absorption observation by S. J. Bell~Burnell \&
W. M. Goss (priv. comm) which implies a distance of $11.5 \pm 1\ \rm kpc$ 
based on $R_0=8.5\ \rm kpc$; and, a Chandra observation of \cyg 's X-ray
scattering halo from which a geometric distance of
$9 \pm_2^4\ \rm kpc$ was calculated \citep{Predehl00}.

Here we present high resolution images of \cyg\ taken with 
the Very Long Baseline Array (VLBA)\footnote{The VLBA is a facility of the
National Radio Astronomy Observatory which is operated by
Associated Universities, Inc. under cooperative agreement with
the National Science Foundation}, following a large radio
flare in February 1997, showing a one-sided radio jet directed south from a
highly-variable core.  \S 2 details the observations and data reduction,
while \S 3 describes the images and model fits, and \S 4 the conclusions.

\begin{figure}[t!]
\plotone{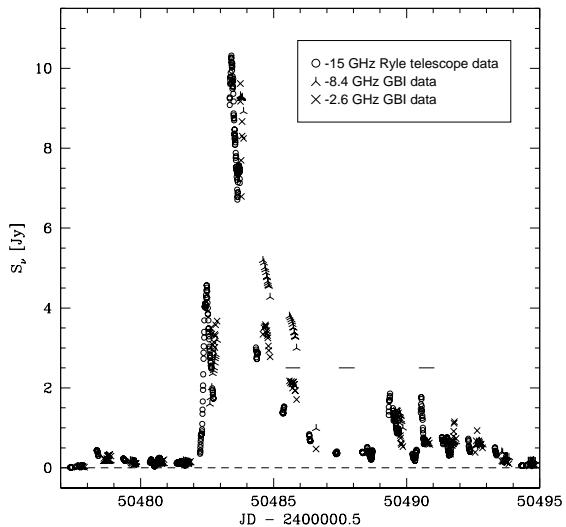}
\caption{Radio light-curves for \cyg\ at the time of our VLBA
  observations.
  The horizontal lines show the times spanned by our VLBA
  observations.  Note that the source was optically thin during the first
  epoch, and optically thick towards the end of the third; there are no
  simultaneous flux measurements for the second epoch.  The x-axis is
  Modified Julian Date: MJD= JD$-2400000.5$.}
\end{figure}

\section{Observations and Data Reduction}

\cyg\ was observed with all 10 antennas of the VLBA in three 13~hour
sessions, two, four, and seven days after the source flared to above 10~Jy
at 15~GHz on
4~February~1997.  Figure~1 shows the epochs of the VLBA observations in the
context of the radio light-curves, taken from the Green Bank Interferometer
(GBI) and the Ryle Telescope (to be discussed in a later paper).  During
most of the first epoch (6 February) observations the source as a whole was
optically thin ($\alpha\sim-0.6$, $S_\nu\propto\nu^\alpha$) and rapidly
decaying, from $2.1$ to $1.75\,\rm Jy$ at 8.4~GHz.  Unfortunately there were
no total flux density measurements during the second epoch (8 February); by
the third (11 February) \cyg\ had begun a series of smaller flares with quite 
rapid declines, the flux density at 15~GHz varying by a factor three or more 
during the VLBA observations.

\begin{figure*}[ht!]
\epsscale{2.0}
\plotone{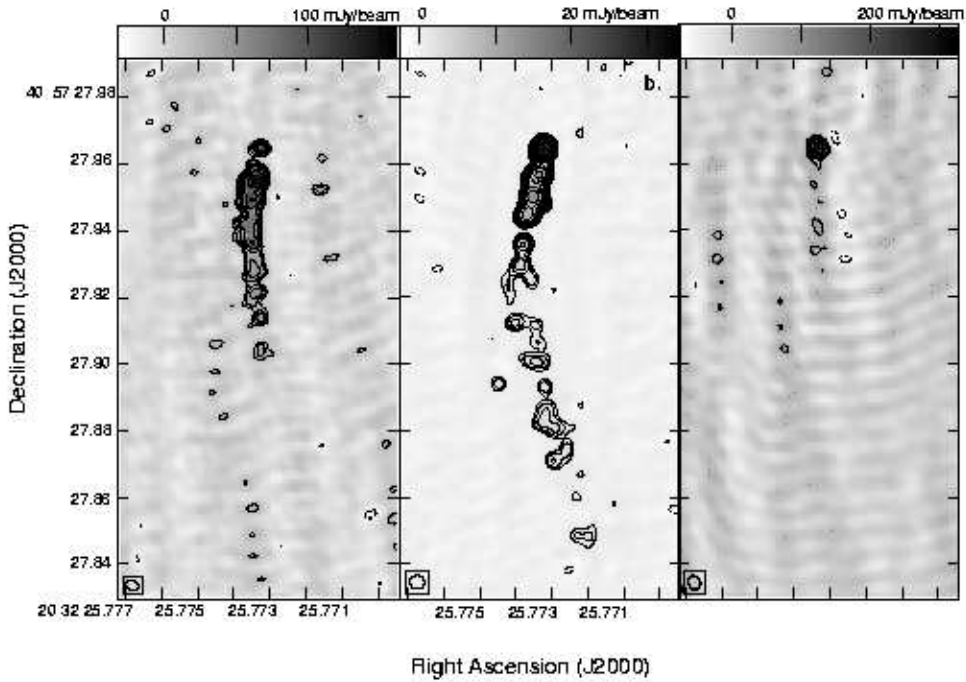}
\caption{VLBA images of \cyg, from 6, 8, and 11 February 1997.
  The restoring beam (a Gaussian fit to the dirty beam) is shown in the lower
  left corner of each image.
a.~6 February 1997. Contours are $\pm20\times2^{n/2}\rm\,mJy/beam$,
    $n=0, 1, 2, \dots$; the peak in the image is $161\,\rm mJy/beam$,
    with the northern core roughly half that bright.  The gray scale
    spans the range of the image, from $-29.4$ to $160.8\rm\, mJy/beam$.
    The rms noise is {$7.5\rm\,mJy/beam$}, and the restoring beam is
    $3.98\times3.00\,\rm mas$.
b.~8 February 1997. Contours are $\pm1.25\times2^{n/2}\rm\,mJy/beam$,
    $n=0,1,2,\dots$; the peak in the image is $33.9\,\rm mJy/beam$.
    The gray scale spans the range of the image, from $-2.0$ to
    $33.9\rm\, mJy/beam$.
    The rms noise is $0.48\rm\,mJy/beam$, and the restoring beam is
    $4.66\times4.13\,\rm mas$.
    As discussed in the text, the time-variable core has been removed and
    replaced by a Gaussian with its mean flux density.
c.~11 February 1997. Contours are
     $\pm 60\times2^{n/2}\rm\,mJy/beam$,
    $n=0,1,2,\dots$; the peak in the image is
    $322.4\,\rm mJy/beam$.
    The gray scale spans the range of the image, from $-73.5$ to
    $322.4\rm\, mJy/beam$.
    The rms noise is $20\rm\,mJy/beam$, and the restoring beam is
    $4.29\times3.81\,\rm mas$.
    Unlike the 8 February image, no attempt has been made to remove the
    time-variability of the core in this image.}
\epsscale{1.0}
\end{figure*}

  A third and potentially more serious complication is the 
motion of the jet during the observations.  As discussed below, the 
proper motion of the jet is $\gtrsim20\,\rm mas/day$, corresponding to
moving by two beams or more during our $\sim 1/2\rm\, day$ observations. 

  The VLBA observations were carried out with a 16~MHz bandwidth centered on 
15.365~GHz, using dual polarization and two-bit sampling.
% This frequency was chosen to minimize source
% broadening by interstellar scattering (Wilkinson, Narayan, \& Spencer 1994;
% Molnar \eta\ 1995). 
The data were correlated using the VLBA correlator in 
Socorro, NM, USA, and amplitude calibration and fringe fitting were performed  using NRAO's 
Astronomical Image Processing System (AIPS).
Scans on \cyg\ were ``sandwiched'' between scans on the calibrator J2025+334,
$7^\circ$ away, with a three-minute cycle time, allowing the use of antenna
amplitude and phase gain solutions from that source as a first estimate for
those on \cyg.

  There were several additional complications to the data reduction. 
First,
  J2025+334 was too far away from \cyg\ to track the latter's phase variations 
  accurately; although the position of the northern-most component (presumably 
  the core) was stable to within a few milliarcseconds, self-calibration was
  required to bring out the full structure of the jet.
Second,
  both \cyg\ and J2025+334 are scatter-broadened, as indicated
  by the fall-off in the interferometric amplitudes on the longest baselines 
  even when the core completely dominates the image (as in the third epoch).
  This may of course be intrinsic source structure, but is consistent with
  the scattering sizes previously measured (see \S3.1). 
  In general we concentrate on
  the more extended jet emission, and the images of all three epochs, shown in
  Figure~2, are made with a $30\rm
  M\lambda$ full-width at half-maximum (FWHM) Gaussian taper.  The 
  corresponding synthesized beams are roughly 3--5~mas FWHM.

Simulations showed that proper motions of this order would have only minor
effects on the images, and would not affect any of our major conclusions.
In particular, we modeled the source as a stationary core with a straight jet
60~mas in length beginning 10~mas away, oriented along a position angle of
$175^\circ$.  The total flux density in the jet was taken to be about 10 times
that of the core, and fell off as $r^{-0.5}$, where $r$ is the distance
along the jet; this roughly matches the image from the first epoch
(Figure~2a).  The simulated jet was taken to move outward as a whole
along the same position angle at 20~mas/day.  The core, and the jet
perpendicular to the ejection axis, were modeled as unresolved, but
convolved with a Gaussian of 2~mas FWHM to simulate the effect of interstellar
scattering.  The Fourier transform of this model was then
projected onto the observed baselines in the uv-plane, and the entire
13~hour data set imaged and deconvolved following the same procedure as used
for the observed data.  The resulting image showed a
slightly elongated version of the model as it appeared at the mid-point of the
observations, with the elongation corresponding to the proper motion of the
jet over 13~hours; this seems intuitively reasonable, and shows that
proper motion does not significantly affect our conclusions.  Following individual radio components between epochs is not advisable,
but the apparent curvature and rough extent are real.  

  The February 6 data were further compromised by poor {\it a priori}
amplitude calibration of several antennae due to snow over most of the
southwestern United States.  This epoch required correspondingly more
drastic self-calibration and careful imaging, and the resulting map
(Figure~2a) is still not nearly as good as those from the second epoch. 

  The 8 and 11 February data were taken in much better weather, and for
these epochs the major complication is the intrinsic variability of the
source.  This produces artifacts which dominate the noise in the 
maps, and makes imaging even the core itself quite difficult.  During the second
epoch (8 February) most of the variability, as well as the total flux, was concentrated in
the core.  We therefore created a core-only data set by subtracting a first
(crude) model of the jet, self-calibrated that data set in hour-long
segments, transferred the antenna gains thus derived to the original data,
and subtracted CLEAN-component models \citep{H74} of the time-variable core.  This
produced a jet-only data set phase-referenced to the core.  Figure~2b shows
the image produced from this data set with one round of phase
self-calibration applied and a Gaussian model of the core with an average
flux added.  The improved calibration reduced the rms noise in the image by
a factor of 5, showing that the jet extends an additional 90~mas beyond
what could be seen in the maps which were phase-referenced solely to
J2025+334.

  Unfortunately this method did not improve the 6 and 11 February data sets.
For 6 February this could be because of the poor initial amplitude
calibration, or because the flux variability was primarily in the jet rather
than in the core; most likely, both effects contribute.
The 11 February map (Figure~2c) by contrast is entirely dominated by
the core, but the above procedure while reducing the noise level did not
reveal any convincing jet, to a limit of 10--20~mJy/beam in a 5~mas beam.
This noise level is far higher than that in the 8 February map (Figure~2b),
presumably
because the core is much stronger and its flux variations far greater (a
factor 3--4 during these observations; see Figure~1).  With this noise level
one could not expect to see any emission corresponding to the 8 February
jet, even if that emission had not faded at all. 

\section{Discussion}

\subsection{Is the Northern Component the Core?}
  One of the primary conclusions of this work is that the jet in \cyg\ is
one-sided, making this the first severely asymmetric Galactic relativistic
source.  This conclusion rests on the premise that the northernmost component
is associated with the core.
The evidence for this is quite compelling.  Although phase-referencing was not
entirely successful, the maps before self-calibration did show that the
brightest (northernmost) component is stationary to within 3~mas.
Further, the strong variations seen with the Ryle telescope on 11 February      are mirrored exactly in the amplitudes measured on the shortest VLBA baselines,
consistent with the corresponding image (Figure~3), which shows no evidence 
for any but
the compact northern component.  This variability, on timescales of minutes
to hours, is most easily explained by assuming this component is indeed the
central core, with the variations caused by short-timescale ejection and
decay of unresolved jet segments.   Finally, simultaneous measurements
with the GBI towards the end of the VLBA observations show that the source
was optically thick ($\alpha\sim0$), a characteristic of emission on very
small scales; the fact that the emission at this time came solely from the
northern component is another argument that that component is indeed the
core.

\begin{figure}[t!]
\plotone{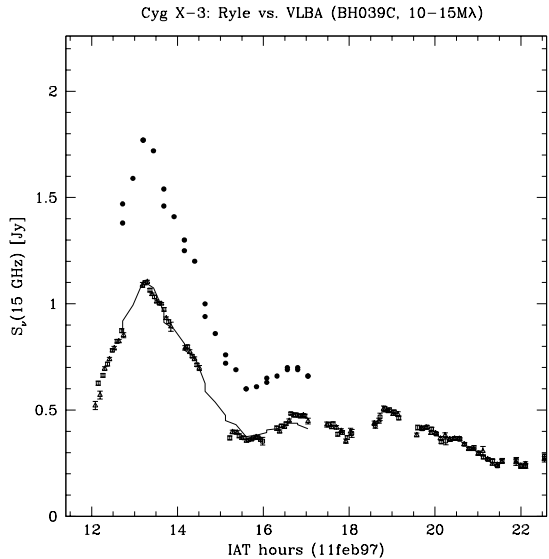}
\caption{Comparison of Ryle Telescope and VLBA flux densities at 15~GHz,
  for 11 February 1997.  The solid circles show the Ryle data, the open
  squares (RCP) and triangles (LCP) the average of the VLBA baselines between
  10 and $15\,\rm M\lambda$.  The solid line shows the Ryle flux densities
  divided by 1.6; the agreement is excellent, apart from this constant
  factor.  This multiplicative offset is due partly to scatter-broadening (for
  any reasonable image the flux density increases as the baseline gets
  shorter, and the Ryle data are effectively taken at $0\,\rm M\lambda$),
  and partly to calibration errors (see text).}
\end{figure}

  One possible counter-argument is that the northern component is clearly
resolved, as
seen both in images and more directly in the uv-plane, where the visibility
amplitude drops off rapidly with radius.  This might be either intrinsic
structure or broadening due to interstellar scattering; \cyg\ is one of
the most scatter-broadened sources known (e.g., Wilkinson, Narayan, \&
Spencer 1994; hereafter WNS), and so might be affected even at these high 
frequencies.
To check this quantitatively, we used K.~Desai's private AIPS task
OMFIT to fit half-hour segments of the 11 February uv-data directly, both
with and without simultaneous short time-scale (10--120s) phase
self-calibration. Segmenting the data was required to disentangle the source
size from the extreme flux variability; fitting in the uv-plane is
preferable to more normal image-plane fitting, both because those data are
the observed quantities, and because the poor uv-coverage in such short
periods makes mapping and deconvolution difficult.  In any case the
results from OMFIT were consistent with more standard fitting in
the map plane, and the residuals were reasonable both in the map and in the
uv plane.  The best-fit Gaussian has a FWHM (geometric mean of major and
minor axes) of $1.7\pm0.3\rm\,mas$, an axis ratio of $0.8\pm0.25$, and is
elongated along a position angle of $50\pm20^\circ$ (the error bars reflect both
statistical errors in the individual fits, and the full scatter between the fits
for the half-hour segments).  This and previously published size
determinations are shown in Figure~4.  Our measurement, as well as the 22~GHz observation of \citet{MRG88}, is
consistent with a $\nu^{-2}$
extrapolation from earlier low-frequency ($\leq5\,\rm GHz$) measurements of
the scatter broadening,
% (Wilkinson, Narayan, and Spencer 1994; Molnar \etal\
% 1995; Schalinski \etal\ 1995; and references therein),
but disagrees both with 8.4~GHz observations by
% \citet[hereafter, GKS]{GKS79},
Geldzahler, Kellermann, \& Shaffer (1979; hereafter, GKS)
and with 15~GHz observations by
Newell, Garrett, \& Spencer (1998; hereafter, NGS).
% \citet[hereafter, NGS]{NGS98}.  
The disagreement with GKS (who obtain a size of
$1.3\pm0.2\rm\,mas$) might be dismissed on the grounds that their result was
based on a single short observation with poor sensitivity and only four
useful baselines; their measurement also falls well below any reasonable
extrapolation of the lower frequency data.  NGS on the other hand observed
for 9~hours with the full VLBA, producing a data set quite comparable to
ours, and obtained mean FWHMs of between 1 and 1.8~mas at roughly the same
frequency (15.3~GHz); they also found that the FWHM was strongly correlated
with the flux density on short VLBA baselines.  (At the current stage of
analysis our data do not demand that the source change size, and in any
case show with high confidence that the size at all times lies within the
bounds noted above.)

\begin{figure}[b!]
\plotone{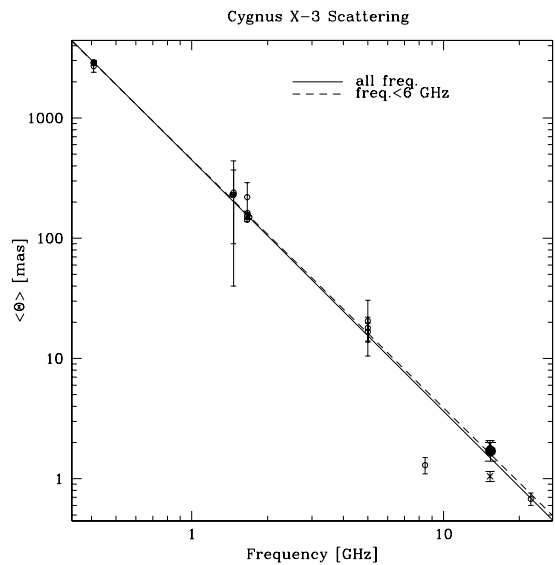}
\caption{The apparent size of \cyg\ as a function of frequency.
  Sizes are geometric means of the minor and major axis FWHM of
  elliptical Gaussian fits, or of circular Gaussians if only those were fit.
  The solid line represents a weighted least-squares fit to all the data,
  $\rm\theta=448\left(\nu/1\,GHz\right)^{-2.09}\,mas$; the dashed line
  gives the fit using only data below 6~GHz,
  $\rm\theta=453\left(\nu/1\,GHz\right)^{-2.07}\,mas$. Data are taken
  from
  \citet{Aetal72};
  \citet{FBGSP95};
  \citet{Getal83};
  GKS;
  \citet{MRG88};
  \citet{MMRJ95};
  NGS;
  \citet{Setal95};
  \citet{SSJH86};
  WNS;
  and this work.  Two values (represented by crosses) are given for
  NGS at 15~GHz, representing the quiescent (minimum) and minor flare
  (maximum) states.  The maximum NGS value is overlaid by our measurement,
  which is indicated by a solid circle.}
\end{figure}

  It is difficult to reconcile NGS' lower bound on the size (1~mas) with the
previous measurements and any reasonable scattering model, since it lies
significantly below a $\nu^{-2}$ extrapolation of those data.  On the other
hand, the size (1.8~mas) and orientation (position angle
$\sim60$--$70^\circ$, based on their images) NGS found when the source
was brighter matches our own.  Further, both the ellipticity and position
angle derived from our data are within the errors identical with those derived
at 1692~MHz
by WNS (1.32 and $62^\circ$, respectively)
and at 1665~MHz by \citet[$1.31\pm0.02$ and
$52\fdg0\pm1\fdg5$]{MMRJ95}.
Admittedly most of the observations, as well as our own, were
taken during or shortly after flares, and NGS found their larger sizes
during short-lived `mini-flares' to 200--300~mJy.  But while NGS find a much
smaller size ($\sim1\,\rm mas$) during quiescence ($S_\nu\sim40$--$60\rm\, mJy$
at 15~GHz),
% \citet{FBGSP95} three-day MERLIN observations at 1658~MHz
Fender et~al.'s (1995) three-day MERLIN observations at 1658~MHz
during another quiescent period (40-50~mJy at both 18 and 2cm)
gave a size which was only slightly smaller than earlier flaring
% measurements ($142\pm5\rm\,mas$, vs. 150--160~mas from \citet{SSJH86},
% WNS, and \citet{MMRJ95} at 1660--1692 MHz) but with a virtually identical
measurements ($142\pm5\rm\,mas$, vs. 150--160~mas from Spencer et~al. 1986,
WNS, and Molnar et~al. 1995 at 1660--1692 MHz) but with a virtually identical
axis ratio ($1.31\pm0.08$) and position angle ($62^\circ\pm3^\circ$).  The
\citet{FBGSP95}\ result seems fairly solid, given the large amount of data 
(three full tracks, yielding an rms noise of $38\rm\,\mu Jy$) and high
resolution (0.15~arcsec).  If the NGS result is correct their observations must
have coincided with a
very unusual period of lower and much more isotropic scattering.
% Anisotropic scattering can readily explain the agreement
% of axis ratios and elongations over such a wide range in frequency and
% angular resolution, while intrinsic structure if large enough to be inferred
% from low frequency data should be immediately obvious at higher frequencies.
% One could invoke a local scattering screen, but this would have to be
% both efficient and stable to produce such a large and consistent effect, and
% would have to disappear during quiescence.  If this were true it would be
% quite exciting; further imaging observations during quiescence are clearly
% needed.
 
  So we are left with two possibilities: either the NGS result for some
reason is spurious, and the northern component is an unresolved source
broadened by interstellar scattering; or the NGS result is correct, and the
northern component we see has roughly the same size and orientation as they
observe
during mini-flares.  Either way it is difficult to imagine that the northern
component is not closely associated with the central object, and it seems
most reasonable to assume that it is the origin of the extensive jet ejected
during the February radio flare.  All the evidence therefore indicates that
our images of \cyg\ indeed show a one-sided jet source.

\subsection{The Jet}
  Accepting the northern component as the core, the jet, shown in Figure
2, is at least 50~mas
long on 6~February, and 120~mas long by 8~February.
These are lower limits on the true length of the jet, as it fades into
the noise towards the south.  In both maps the jet is
curved, and that curvature changes between the two epochs.

  What can we learn from these images?  Most obviously, the ejection
must be fairly continuous, a stream rather than one or a few `blobs.'
This is reminiscent of the jets seen in SS433 \citep{VSIFS87} and
GRO~J1655-40 \citep{HR95}, but unlike the easily-separable
components seen in GRS~1915+105 \citep{MR94, Fender99}.  Unfortunately
this, combined with the poor quality of the 6 February image, makes it
impossible to measure the proper motions of individual components directly.

Based on the radio light curves (Figure 1) it seems likely that the jet was
ejected at the time of the large radio flare, beginning roughly at
MJD 50482.1$\pm$0.1 (MJD -- Modified Julian Date: JD$-$2400000.5).
This is not absolutely
conclusive --- the amount of flux in the most extended structures is not so
great that it would necessarily have dominated the light curve even a week
or more before the flare, if it did not decay over time.  But the jet
emission has clearly declined significantly from the first to the second
VLBA images (Figures 2a and 2b), and this together with the quite sharp
decline seen in most of the smaller radio flares
strongly suggests that the emission we see originated in the main radio
flare on MJD 50482.1.

%
%%%%%
%

\begin{deluxetable}{ccccc}
\tablecolumns{5}
\tabletypesize{\footnotesize}
\tablecaption{Observations}
\tablehead{
  \multicolumn{2}{c}{Epoch} &
  \colhead{time since start of flare} &
  \colhead{proper motion} &
  \colhead{}\\
   MJD  & date  &
  (days) & (mas/day) & \colhead{$\beta_{app}$}}
\startdata
  50485.46--50486.00 & (6~Feb.) & $(3.4-3.9)\pm0.1 $
     & $14.7\pm0.4-12.8\pm0.3$ & $\gtrsim(0.85-0.74)\pm0.02 $ \\
  50487.46--50488.05 & (8~Feb.) & $(5.4-5.9)\pm0.1 $
     & $21.5\pm0.4-19.7\pm0.3$ & $\gtrsim(1.24-1.14)\pm0.02 $ \\
  50490.46--50491.04 & (11~Feb.) & $(8.4-8.9)\pm0.1$ & \nodata & \nodata \\
\enddata
\tablecomments{The proper motions are based on lengths of
  50 mas and 120 mas for 6 and 8~February, respectively.
  The apparent velocity is based on a distance of 10 kpc.
  }
\end{deluxetable}

%
%%%%%
%

Table~1 lists the epochs of observation, inferred ages, proper motions and
apparent jet speeds.  These proper motions ($14-21\rm\,mas/day$) are 
significantly higher than those previously cited for this
source, which range from 8.4 \citep[corrected for a
one-sided jet]{Setal95} to $10\pm2$ \citep{Getal83} to 4.6--18~mas/day
\citep[with the range corresponding to the uncertainty in the
ejection date]{SSJH86}.  Given the limitations of those earlier observations, which
could easily have missed the sort of low-level extended structure on which
we base our higher estimate, there is no evidence that the proper motion has
changed with time.  To the contrary, the agreement between the position
angles measured in those earlier observations (in all cases almost directly
north-south) and our own suggests that the ejection axis at least has remained
remarkably stable since at least 1982.

  One possible inconsistency is that the uv-plane fits to the data from
the third epoch do not suggest such large source motions.
Probably the explanation is that the emission associated with short timescale
flares is intrinsically different from the longer-lived emission
characteristic
of larger flares like that of 4 February; for instance, it might be that
the emission associated with the short flares decays too rapidly to be seen
outside the scattering disk.  For the remainder of
this section we assume that the third epoch provides no relevant information
on the motions of the jet seen in the first two epochs.

\subsubsection{Physical Parameters}

  Our images give three basic observables useful for deriving the intrinsic
properties of the jet: 
  the apparent proper motion; 
  a limit on the ratio of the brightness of the approaching to that of the 
    receding jet, if the system is intrinsically symmetric;
  and the appearance of the jet itself, in particular its curvature and
    evolution with time. 
The first two are
susceptible to direct mathematical analysis; the last requires more
heuristic modeling, which is deferred to the next subsection.
Note that we implicitly assume that the observed proper motion is related to
a physical velocity, rather than a group speed or some optical illusion
involving e.g. scattering screens or the simultaneous `lighting up' of
well-separated hot spots.

  Under this assumption, according to special relativity
% \citep{HM91},
(e.g., Hughes \& Miller 1991),
a jet moving at an intrinsic
speed $\beta c$ at an angle $i$ to the line-of-sight will be observed to have
an apparent transverse motion $\beta_{\rm app}c$, where
  $$\beta_{\rm app}= {\beta\sin i\over 1-\beta\cos i}$$
The minimum $\beta$ for the approching component, for a given $\beta_{\rm app}$ is
  $$\beta_{\rm min}= {\beta_{\rm app}\over\sqrt{1+\beta_{\rm app}^2}}$$
corresponding to an inclination of
  $\sin^{-1}\sqrt{1\over1+\beta_{\rm app}^2}$.
As discussed above, for our observations
  $\beta_{\rm app}\gtrsim(1.24-1.14)\pm0.02$,
implying
  $\beta_{\rm min}\gtrsim(0.78-0.75)$.
Since $\beta\leq1$, the inclination lies between $0^\circ$ and
$$i_{\rm min}=\sin^{-1}{2\beta_{\rm app}\over1+\beta_{app}^2}$$
For \cyg\ this implies
  an inclination between $0^\circ$ and $(78-83)^\circ\pm1^\circ$.
The corresponding Doppler boosting (see below) is unconstrained;
for inclinations above $59^\circ$, the source could
actually be {\it de-boosted}, i.e. appear fainter than it would without
relativistic effects, as has been seen in GRS~1915+105 \citep{MR94,Fender99}.
%and in extragalactic jet source Cygnus~A  \citep{CBL91}.

  If the source were intrinsically symmetric, the ratio of the flux in the
approaching to that in the receding jet would provide an additional constraint.
Since we do not detect a jet on the opposite side, we have only an upper
limit on this ratio: $R_{\rm obs}\gtrsim 330$, determined by integrating the
flux in the southern jet and comparing that to the noise to the north,
integrated over a somewhat smaller region to account for the slower (apparent)
motion of the receding jet.  The Doppler factor for a
relativistic jet is
  $$\mathfrak D={\sqrt{1-\beta^2}\over1-\beta\cos i}$$
The corresponding Doppler boosting of the observed flux density is \citep{PZ87}
  $${S_\nu\over S^\prime_{\nu^\prime}}=\mathfrak D^{k-\alpha}$$
where $\nu$ is the frequency, $S_\nu$ the flux density, $k$ is a constant
($k=2$ for a continuous jet, $k=3$ for an unresolved blob), and $\alpha$
is the spectral index ($S_\nu\propto\nu^\alpha$).  The primed quantities
refer to the rest frame of the object.  The predicted flux ratio (between
the approaching and receding jets) is then
  $$R=\left(1+\beta\cos i\over1-\beta\cos i\right)^{k-\alpha}$$
where $i$ is now taken as the inclination of the approaching jet.  Solving
for the intrinsic as a function of the observed quantities, we have
$$\beta\cos i={R^{1\over k-\alpha}-1\over R^{1\over k-\alpha}+1}$$
Since both $\beta$ and $\cos i$ can be at most 1.0, $R\geq1$,
and $(k-\alpha)>0$ for synchrotron emission at these frequencies, an
observed lower limit on $R$ corresponds to a lower limit on $\beta$ and an
upper limit on $i$.  The GBI radio light curves during the period when these
observations showed a jet suggest $\alpha\sim -0.6$, typical for
optically-thin synchrotron emission; since the jet appears fairly
continuous, we take $k=2$.  $R\gtrsim330$, as discussed above.  We have then
that
  $$\beta\cos i\gtrsim 0.806$$
implying $\beta\gtrsim 0.81$ and $i\lesssim 36^\circ$.  These limits are
consistent with, and more stringent than, those based on the proper motion.

  Clearly it would be useful to place limits on the maximum as well as the
minimum proper motion.  Each epoch's observations took place over 13~hours;
one could therefore hope to check for changes in source structure within a
given observing run.  Unfortunately the limited uv-coverage
within a small segment makes it virtually impossible to image such a complex
structure as the curved jet; further, the corresponding point spread
function changes drastically as the earth rotates, making even an
unchanging source appear quite differently when imaged with short segments
of data separated by even a half hour. 

\subsubsection{Jet Curvature and Precessing Jet Models}

  The simplest explanation for the curvature of the jet, and the change in
that curvature between the first and the second image, is that the jet is
precessing (cf. Hjellming \& Johnston 1988). The fact that the jet
curves out to the west, then back to the east, with no sign of another
westward bend, implies that the ejection occured over at most one full
precession period.  Given the length of the jet this implies a
joint limit on the speed and the period:
  $$ P\,\left(c\,\beta \sin i\right)\gtrsim l_{jet}$$
with $l_{jet}$ being the length of the jet,
  $6.7$~light-days on 8~February, assuming a distance of 10~kpc.
The precession period $P$ must therefore be
  \begin{eqnarray*}
    P\gtrsim l_{jet}/\left(\beta \sin i\,c\right)\approx
              6.7{\,\rm days}/\beta \sin i
  \end{eqnarray*}
Since $\beta\,\sin i\leq1$,
   $P\gtrsim6.7\,{\rm days}$.

  To proceed further requires more detailed modeling.  Given the complex
radio light curve and the various imaging problems mentioned above,
and to avoid the additional assumptions involved in physically modeling the
ejecta, we attempt to fit only the shape of the jet and its evolution between
the two epochs, not the jet's brightness
distribution.  That is, we only require that the models trace the jet geometry 
in both the first and the second epoch.
We further assume, as in the last
paragraph, that the observed jet was produced within a single precession
period.

  The precessing jet model as described by \citet{HJ88}
has eight important parameters:
  the distance, $d$;
  the precession period, $P$;
  the intrinsic velocity, $\beta c$;
  the inclination angle of the jet axis to the observer, $i_{ja}$;
  the cone opening angle, $\psi$;
  the position angle of the jet on the sky, $P.A.$;
  the phase of the precession at some fiducial time, $\phi_0$;
and the sense of rotation of the jet (clockwise or counter-clockwise, as
seen from the jet origin).  The distance of the source is kept fixed at
10~kpc; the other parameters are to be found by fitting the
observed VLBI images.  In order to find the full range of allowed parameters,
we searched for the best alignment between the images
and the model, based on $\chi^2$ minimization \citep{PFTV86}, using several million
randomly-chosen sets of input parameters.  The initial guesses uniformly
covered the full range of physically meaningful model parameters (see
Table~2, with the maximum period (600~days) chosen as an
approximation of `very long'.
% Inclinations larger than $90^\circ$ correspond to one-sided jets pointing
% {\it away} from the observer.
The agreement between the
model and the data was measured by the weighted sum of the squares of
the distances between the closest points in the model, and $\sim10$~positions
measured along the `spine' of the jet in each of the first and second epoch
images (see Figure 5a).  The weights were taken from rough error bars based on the 
believability of a feature and the approximate accuracy of its position, 
taking into account uncertainties due to noise in the images, deconvolution 
artifacts, the local width of the jet, etc.  While the absolute value of the 
resulting $\chi^2$ is not very meaningful, models with lower $\chi^2$ do match
the images better, which is all that is necessary for the minimization.

\begin{figure*}[t]
\epsscale{2.0}
\plotone{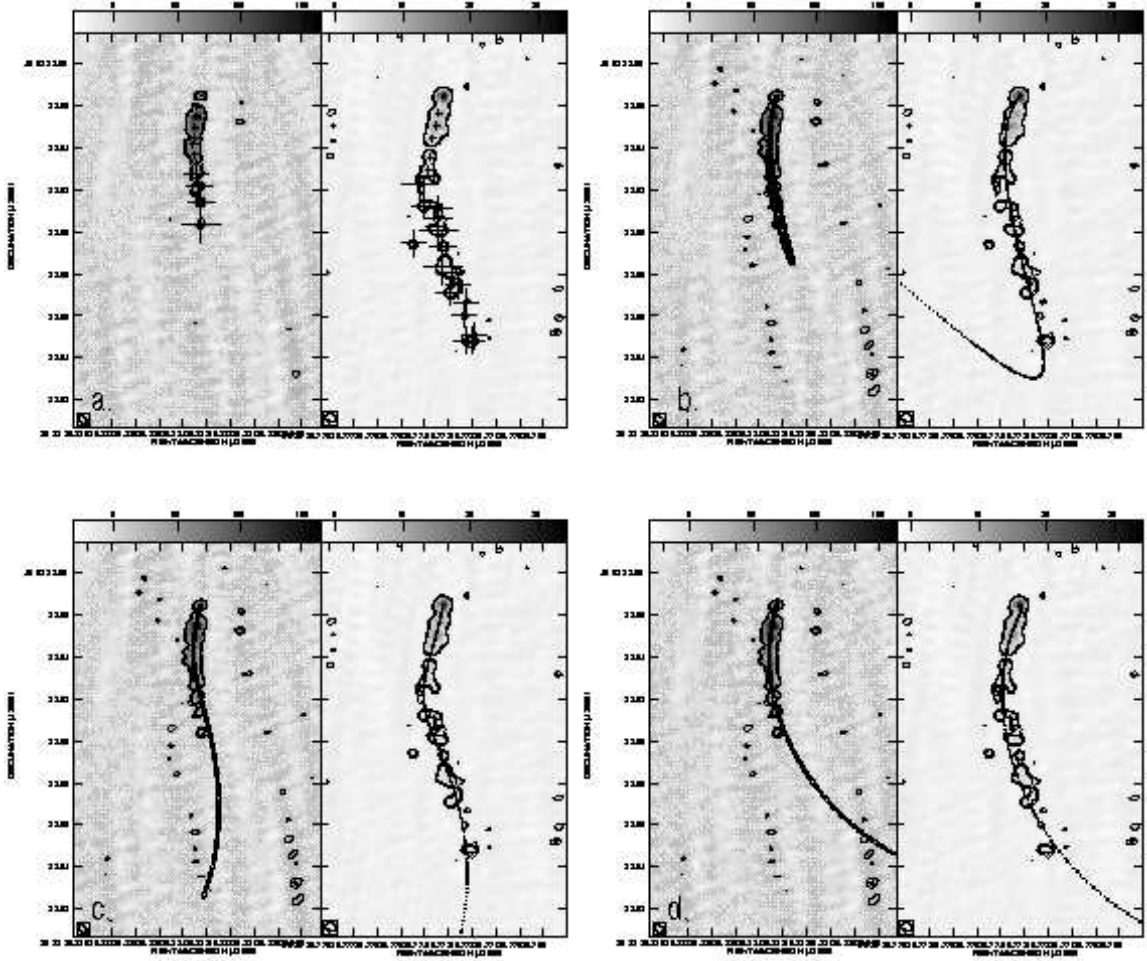}
\caption{Images of 6 February and 8 February observations with,
  contour levels of  $0.025 \times (1, 1.4)\ \rm mJy/beam$ for the 6 February image
  and $0.0012 \times (1, 1.4)\ \rm mJy/beam$ for the 8 February image, and
  a. the positions of the points used for in the model fits superimposed.
     The size of the crosses are $3 \sigma$ and show the relative weight of
     each point (i.e., the smaller crosses have more weight in the model fit).
  b. a jet model with parameters of $P=29.8$~days, $\beta=0.990$, $i_{ja}=82\fdg1$, $\psi=55\fdg7$,
    counter-clockwise rotation; $\chi^2=6.95$ superimposed.
    This model corresponds to the high-inclination branch
    of fits satisfying the age constraint ($\rm MJD_{ej}\geq50482.0$).
    The corresponding peak Doppler ratio $R$ is only 2.3 however, so these
    fits would require an intrinsic jet asymmetry or obscuration to explain
    the lack of a counter-jet.
  c. a jet model with parameters of $P=116$~days, $\beta=0.902$, $i_{ja}=6\fdg3$, $\psi=1\fdg1$,
    clockwise rotation; $\chi^2=7.99$ superimposed.
    A barely-acceptable fit, this is an example of the clockwise-rotating
    family of solutions which satisfies the boosting but not the age
    constraint (it requires a minimum jet age of 7.1~days on 8~February).
  d. a jet model with parameters of $P=496$~days, $\beta=0.967$, $i_{ja}=1\fdg5$, $\psi=1\fdg2$,
    counter-clockwise rotation; $\chi^2=6.44$ superimposed.
    The best-fitting model of the family which satisfies all three
    ($\chi^2$, boosting, and age) constraints.}
\epsscale{1.0}
\end{figure*}

  The initial parameters, and the positions and error bars
measured off the two images, were passed to a $\chi^2$-minimization routine
based on the downhill simplex method \citep{PFTV86},
% About ten million initial, random guesses were used, yielding
producing about a million converged
solutions.  Based on those results several hundred thousand more models
were run, using a more restricted set of inputs chosen to maximize the
number of `good' solutions (defined below), in order to flesh out the
range of acceptable model parameters.
We have three constraints on these solutions.  
First, they must match the observed jet morphology.  Empirically
  $\chi^2\lesssim8$ corresponds to a good chi-by-eye fit.
Second, if the extended jet did indeed originate at MJD 50482.1$\pm$0.1,
  the jet as seen in the first image can be at most 4.0~days old (as seen by
  us), while the jet seen in the second image can be at most 6.0~days old.
Finally, if the jet is intrinsically two-sided, the jet/counter-jet ratio
  on 8~February is $\gtrsim330$ (see \S 3.2.1).  Without a model for the jet
  brightness distribution and its evolution, this last constraint is
  impossible to apply fully to these models.  We take a conservative
  approach, requiring that at least one observed point along the jet
  have a boosting factor, relative to the corresponding component in the
  purported counter-jet, of at least $330^{1/2.6}=9.3$ (see \S 3.2.1). 
Figure 5b-c shows the images with examples of the major families of 
solutions superimposed.
% The results of
% applying these three constraints to the solutions derived from the
% $\chi^2$-minimization routine are given in Table~2.

  The results are given in Table~2.
  Good fits, as measured by $\chi^2$, may be obtained for a wide range of 
parameters (see Figures 6 \& 7). In particular, jets pointing away from the observer can match 
the observed morphologies in both images.  There is a lower limit on the 
period of about 15~days, analogous to the analytic limit discussed above.  
More surprisingly, there is also a lower limit on the age of the ejecta: they 
must have been expelled at least 2.5 days before the first image, and 
at least 3.5 days before the second.  Note that this is derived {\it simply 
by fitting the morphology}, without any additional constraints from the 
radio light curves, and results from the requirement that both images be fit 
simultaneously
within one period, which limits the proper motion and therefore the age.
Interestingly, this lower bound on the age corresponds to the
peak in the radio light curves around MJD 50483.5.

\begin{figure}[ht!]
\plotone{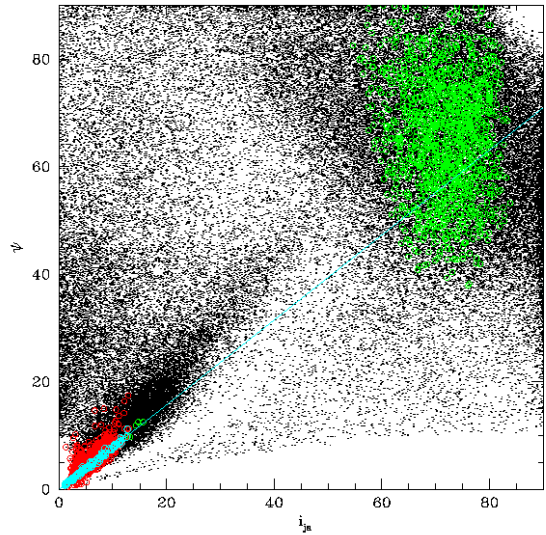}
\caption{Inclination of the jet axis, $i_{ja}$, vs. cone opening angle,
  $\psi$, for `good' model fits to the 6 and 8~February 1997 images.
  For simplicity, only those fits with counter-clockwise jet rotation and
  $i\leq90^\circ$ are shown.
  Black dots represent fits with $\chi^2\leq8$.  Requiring in addition
  that jet ejection begin no earlier than MJD 50482.0 (see text) restricts
  possible fits to those indicated by the green (and blue) circles.
  Similarly, requiring that the jet's apparent asymmetry be due to Doppler
  boosting results in the possible fits shown by the red (and blue) circles.
  Imposing
  both age \& asymmetry conditions at once gives the single-parameter family
  of solutions shown as the blue circles.  A least-squares fits to these
  latter points gives the blue line:
  $\psi\approx 0.79\,i_{ja}-0.08^\circ$.}
\end{figure}

%
%%%%%
%

\begin{deluxetable}{lccccccc}
\tablecolumns{8}
% \rotate
\tabletypesize{\small}
\tablecaption{Results of Modeling Jet Morphology}
\tablehead{
 \colhead{} & 
   \colhead{      } &
   \colhead{Period} & \colhead{       } & \colhead{        } &
   \colhead{      } & \colhead{      }  & \colhead{        } \\
             &
    \colhead{Rotation} &
    \colhead{(days)} & \colhead{$\beta$} & \colhead{$i_{ja}$\tablenotemark{a}} &
    \colhead{$\psi$} & \colhead{$P.A.$}  & \colhead{$\phi_0$} \\
  }
\startdata                                                                     
Initial guess\tablenotemark{b} &
  CW, CCW\tablenotemark{c} &
  $0-600$\tablenotemark{d} & $0-1$   & $0-180^\circ$ & $0-90^\circ$  &
  $0-360^\circ$ & $0-360^\circ$ \\
\\
%
%   P         &  beta   &    i    &   psi   &   PA    &  phi0 
$\chi^2\leq8$ &
  \nodata     &
  $\gtrsim10$ & \nodata & \nodata & \nodata & \nodata & \nodata \\
\\
$\chi^2\leq8$ \&\ $\rm MJD_{ej}\geq50482.0$\tablenotemark{e} &
CCW &
$>50$   & $>0.75$ & $<18^\circ$ & $<15^\circ$ &
$120-160^\circ$ & $50-100^\circ$ \\
&CCW &
  $20-60$ & $>0.75$ & $50-85^\circ$& $35-90^\circ$ &
  $210-275^\circ$ & $265-325^\circ$ \\
\\
$\chi^2\leq8$ \&\ $R\gtrsim330$ &
  CCW &
  $>60$   & $>0.81$ & $<13^\circ$ & $<18^\circ$ &
  $60-300^\circ$ & $>50^\circ$ \\
 &CW &
  $>70$   & $>0.81$ & $<15^\circ$ & $<10^\circ$ &
  $180-220^\circ$ & $135-270^\circ$ \\
\\
$\chi^2\leq8$, $\rm MJD_{ej}\geq50482.0$\tablenotemark{e}, & & & & & & &\\
\quad \&\ $R\gtrsim330$ &
CCW &
  $>60$   & $>0.81$ & $<13^\circ$ & $<12^\circ$ &
  $125-155^\circ$ & $50-100^\circ$ \\
\enddata

\tablenotetext{a}{Inclinations larger than $90^\circ$ correspond to one-sided 
jets pointing {\it away} from the observer.}
\tablenotetext{b}{Initial guesses were taken as uniform random deviates over
the listed range, with each parameter chosen independently.}
\tablenotetext{c}{Both clockwise (as seen for the jet origin) and
counter-clockwise jet rotation were allowed.}
\tablenotetext{d}{The maximum period (600~days) was chosen as an
approximation of ``very long.''}
\tablenotetext{e}{$\rm MJD_{ej}$ is the date at which jet ejection began.}
\end{deluxetable}

  Requiring in addition that the jet in the first image be no more than 4.0
(and the second no more than 6.0) days old, restricts the possible
solutions enormously.
Jets pointing away from the observer are entirely eliminated,
  because they cannot produce a long enough jet in the required time.
Solutions with clockwise rotation (as seen from the jet origin) are also
  excluded, as they give poor fits to the observed morphology
  (minimum $\chi^2=8.5$).
% determined {\it after} the choice of the cut-off
% $\chi^2=8$).
The solutions in which the jet rotates counter-clockwise are themselves split
into two basic families: 
  low-inclination jets ($i_{ja}\lesssim20^\circ$) with a linear relationship
    between inclination and cone opening angle, and relatively long periods
    ($P\gtrsim50\rm\,days$); and
  high-inclination jets ($i_{ja}\gtrsim50^\circ$) with large cone opening 
    angles ($35-90^\circ$) and relatively short periods.
The latter branch is eliminated if sufficient boosting is required to 
  conceal the counter-jet in an intrinsically symmetric system. 
The solutions obtained using all three constraints then form a one-parameter
family, with
\begin{eqnarray*}
{\rm rotation} & {\rm counter-clockwise} &\\
     P & \gtrsim  60\,{\rm days} &\\
  \beta& \approx \left(-11.89/P\right) + 0.989&\gtrsim0.81\\
i_{ja} & \approx \left(785\fdg5/P\right) -0.25^\circ& \lesssim 14^\circ\\
   \psi& \approx  0.79\,i_{ja}-0.08^\circ  & \lesssim 12^\circ \\
   P.A.& \approx  145\pm5^\circ&\\
 \phi_0& \approx  70\pm10^\circ&\\
\end{eqnarray*}
With these additional constraints, the minimum age for the first image
is $\sim2.8\,\rm days$, corresponding to ejection on
MJD~50482.7--50483.2. 
The limit on $\beta$ arises directly from the jet/counter-jet ratio;
the maximum ages for the two images constrain $\beta_{app}$, which is a
  combination of $\beta$ and $i$; 
those in turn constrain the period, assuming the jet seen in the image was 
  formed within a single precession period.  
The requirement to match the observed curvatures reduces this two-dimensional 
  space of solutions to a single dimension, and further requires the above 
  restrictions on the cone opening angle $\psi$ and the other geometric 
  properties ($P.A.$, $\phi_0$).

\begin{figure*}[t]
\epsscale{2.0}
\plotone{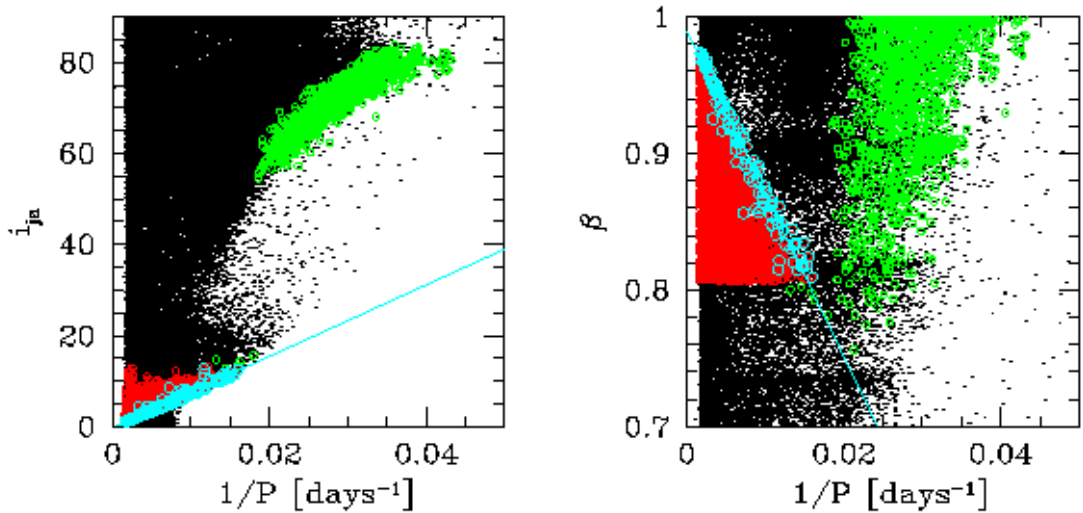}
\caption{Inclination of the jet axis, $i_{ja}$ ({\it left}), and
  intrinsic jet speed, $\beta$ ({\it right}), as a function of the inverse
  period, $1/P$.  Colors are as in Figure~6.  The blue lines
  indicate least-squares fits to the blue circles:
  $i_{ja} \approx \left(785\fdg5/P\right) -0.25^\circ$;
  $\beta  \approx \left(-11.89/P\right) + 0.989$.
  The gap on the left-hand side of each plot corresponds to the maximum
  period allowed as an initial guess, 600~days.}
\epsscale{1.0}
\end{figure*}

  In sum, the simple precessing-jet model of \citet{HJ88}\ can 
fit the observed jet quite well, for either a one- or a two-sided jet.
To conceal the counter-jet requires substantial Doppler boosting ---
% but since
% the cone angle for these boosted solutions is so small, the range in Doppler
% boosts over the period of the jet is not large (at most 70\%), and such
% differential boosting is unlikely to explain short time-scale variations in
% the radio light curves.
in these models the intrinsic (as compared to the differential
jet/counter-jet) Doppler boosting of the flux is at least a factor 13, and
could be much higher.  Also, for these low-inclination models, temporal
variations in the jet's frame are significantly compressed in ours because
of time dilation (e.g.,
in one model 6 months of jet evolution looks like only 6 days to the 
observer of the jet).
This might
help explain both the unusual strength and the rapid variability of Cygnus~X-3
as a radio source, although it should be noted that this time dilation would not
affect variability that originates in the accretion disk or  any other part
of the system that
is not moving towards us at relativistic speeds.
The small range of position angles observed over the years is also consistent
with the small range of ejection angles in these models -- jet models more
nearly in the plane of the sky require much larger cone angles to fit our
images, giving correspondingly wider position angle swings over the full
precession period.

  One argument against the low-inclination models is that the
periodic dips in the X-ray and infrared light-curves are usually modeled as
opacity effects in a reasonably edge-on disk (e.g., Mason, C\'ordova, \&
White 1986).  If the radio jet
were perpendicular to the disk this would suggest the jet must be in the
plane of the sky.
However,
\citet{GEWS81} showed that the X-ray light curves could be matched
in detail even for disks with inclinations up to $70^\circ$, and observations
at other wavelengths are also consistent with a wide range of inclinations
\citep{SGS96,HSF00}.  Also, we clearly see
precession in the jet, which implies that the jet is not
perpendicular to the binary system's plane of orbital motion.  So our solutions
do not put a tight constraint on the inclination of the binary system.

  It is also possible that the system is edge-on and the jet in \cyg\ is in
the plane of the sky.  The one-sided jet could then be explained either as 
intrinsic or, as suggested by \citet{FHP99}, by obscuration.
% of one side of the jet.
Obscuration seems very unlikely, since it would require not only
an odd geometry, blocking out one side of the jet but not the
core or the other side of the jet, but also material opaque at 15~GHz out to
$\gtrsim1200\rm\,AU$. 

\section{Conclusions and Implications}  
  
  The main result of these observations is to show that the radio emission
from \cyg\ during flares is dominated by a one-sided relativistic jet with
an intrinsic speed of at least $0.81c$.  Assuming the jet to be intrinsically
symmetric, precessing jet models give a maximum inclination to the line-of-sight of
$\sim14^\circ$.  \cyg\ is the most luminous X-ray binary at radio
wavelengths, and the observed asymmetry and high proper motion make it
tempting to speculate, as mentioned in the previous section,  that this high luminosity is due to significant
Doppler boosting of a jet pointed almost directly towards us.  This might
also help explain the rapid and extreme radio variability of this and other
known relativistic jet sources. However, since no counterjet is seen,
the VLBA observations do not {\it require} such boosting, and, unless the
X-rays are also associated with the jet (contrary to most current models),
boosting of the jet would not explain why \cyg\ is also one of the brightest 
X-ray sources in the Galaxy ($10^{37}-10^{38}\rm\,erg/s$, Bonnet-Bidaud \&
van~der~Klis 1981).

  There are three other relativistic jet sources in the Galaxy which have
been studied in some detail: SS433, GRS~1915+105
and GRO~J1655-40. \cyg\ is the only one-sided jet
among the four, presumably because at the time of the flare it was aligned more nearly along the
line-of-sight, although the jet GRO~J1655-40 is sometimes
intrinsically asymmetric
\citep{HR95}.  Like GRO~J1655-40, \cyg\ is a strong black hole
candidate based on its mass function, a conclusion strengthened by the limit
derived here on the inclination of the jet, which is probably roughly
aligned with the angular momentum axis of the orbit.   \cyg\ is however the
first consistently strong X-ray source to exhibit such highly relativistic
jets; GRS~1915+105 and GRO~J1655-40 are both X-ray transients, often
undetectable but occasionally among the brightest sources in the sky, while
SS433 is at best an undistinguished X-ray source.
Similarly, only \cyg\ and SS433 are detectable in the radio even when they
are not 
flaring, though GRS~1915+105 can remain in a plateau state for 
months or longer \citep{Fetal96}.  The radio and X-ray states are closely coupled
for all three of the highly relativistic jets, while for SS433 the lack of
an obvious connection may simply be due to the relatively poor X-ray
coverage.  Intriguingly, both GRO~J1655-40 and GRS~1915+105 have a very
unusual, hard power-law X-ray tail to energies of several hundred keV, while
\cyg\ has become famous as an occasional source of these \citep{MFBGS96}
and even higher energy photons (certainly up to 100~MeV; see
% \citet{L77} and \citet{FTL87}).
Lamb et~al. 1977 and Fichtel, Thompson, \& Lamb 1987).

  The clear implication is that changes in the accretion disk produce
changes in the radio jet, and that an excess of high-energy photons may
indicate a source capable of producing highly relativistic radio jets.
It is not particularly surprising that the 
high-energy photons and the high-energy electrons should be fairly closely 
coupled.  What is new here is that an X-ray binary, with consistent and 
reasonably strong X-ray and radio emission --- and hence, a relatively stable
accretion rate, and accretion disk --- can give rise to the highly 
relativistic jets previously associated only with very unusual X-ray
transients. 

  An obvious question is whether the quiescent radio emission is
also in the form of a relativistic jet.  In SS433, it clearly is, and
neither the intrinsic speed nor the orientation of the jet depend on the 
strength of the X-ray or radio emission.  Neither of the highly relativistic
X-ray transients has been detected in radio quiescence, although
observations during the smaller flares  of GRS~1915+105 are broadly
consistent with the long-lived flare imaged by \citet{MR94}.
Unfortunately the current data on \cyg\ are still too
fragmentary to convince one way or the other.  Our 11 February observations
suggest at most a much {\it slower} expansion rate than during giant flares,
while NGS find something much {\it faster}.  Despite this confusion \cyg\
offers the unique opportunity to check the behavior of a highly relativistic
jet source over its whole range of X-ray and radio states; we and doubtless
others will be observing this source for some time to come.

  Finally, with a convincing jet found in an X-ray binary like \cyg, it
is beginning to seem that every Galactic X-ray source with radio emission 
turns out, when imaged, to be a relativistic jet.  While the four discussed so 
far are the only truly compelling examples, there are hints that several 
others are jets as well (possible polarization in 4U 1630-47,
% \citet{buxton98};
Buxton et~al. 1998; elongated 
structure in GX~339-4,
\citet{Fetal97};
suggestions of elongation and 
high-speed expansion in LSI+$61^\circ303$,
% \citet{PGT98};
Peracaula, Gabuzda, \& Taylor 1998;
VLA images of one-sided jet in V4641 Sgr,
% \citep{hjellming00}).
Hjellming et~al. 2000). 
On the other hand, all the relativistic sources currently known are 
very peculiar in some ways, although few of those peculiarities are the same 
for all four sources.  The next challenge is to image some of the fainter,
more common radio X-ray binaries, to see whether more `normal' systems also
give rise to relativistic jets.

\acknowledgments

This project could not have been done without the Green Bank Interferometer
(GBI), which provided the flux measurements which triggered our
target-of-opportunity observations.  Those observations themselves could not
have succeeded without the eager help of many people associated with the
VLBA, particularly the site techs who ran up various mountains to change
tapes at very odd hours; we are most grateful both to them and to the scheduling
committee for their timely efforts.   M.R. enjoyed a month's hospitality at
the RCfTA at Sydney University.  Finally, Ketan Desai kindly provided,
discussed, and improved his uv-plane fitting program in response to our
requests.
The GBI is a facility
of the National Science Foundation operated by the National Radio Astronomy
Observatory (NRAO) in support of NASA High Energy Astrophysics programs.
NRAO is a facility of the National Science Foundation operated under 
cooperative agreement by Associated Universities, Inc.  
A.M. acknowledges support from the European Commission's 
TMR/LSF Programme (Contract No. ERB-FMGE-CT95-0012). 
Basic research
in radio astronomy at the Naval Research Laboratory is funded
by the Office of Naval Research.
The Ryle Telescope is supported by PPARC.

\newpage

%%%%%%%%%%
%

\end{document}